\documentclass[conference]{IEEEtran}
\IEEEoverridecommandlockouts

\usepackage[unicode]{hyperref}

\hypersetup{
	pdfstartview=FitH,
	bookmarksnumbered=true,
	bookmarksopen=true,
	colorlinks,
	linkcolor={blue!90!black},
	citecolor={blue!60!black},
	urlcolor={blue!80!black}
}

\usepackage{multirow}
\usepackage{amsmath,amsthm,amsfonts} 

\usepackage{amssymb}
\usepackage{framed}
\usepackage{soul}
\usepackage{enumitem}   
\usepackage{balance}

\newcommand{\oursname}{Orchestra}

\usepackage{mathtools}

\theoremstyle{plain}

\theoremstyle{definition}

\theoremstyle{remark}

\usepackage{bm}
\usepackage{graphicx}
\usepackage{textcomp}
\usepackage{xcolor}
\usepackage{subfigure}

\usepackage{epstopdf}
\usepackage{balance}

\usepackage[table]{xcolor}

\usepackage{listings}

\usepackage{cite}

\begin{document}

\title{Accurate Table Question Answering with Accessible LLMs} 
\author{
\IEEEauthorblockN{
Yangfan Jiang\IEEEauthorrefmark{0\text{1}*}\thanks{*Work done as an intern at Tongyi Lab, Alibaba Group.},
Fei Wei\IEEEauthorrefmark{0\text{2}}, 
Ergute Bao\IEEEauthorrefmark{0\text{3}}, 
Yaliang Li\IEEEauthorrefmark{0\text{2}}, 
Bolin Ding\IEEEauthorrefmark{0\text{2}},
Yin Yang\IEEEauthorrefmark{0\text{4}},
and Xiaokui Xiao\IEEEauthorrefmark{0\text{1}}
}
\IEEEauthorblockA{\IEEEauthorrefmark{0\text{1}}National University of Singapore;    \IEEEauthorrefmark{0\text{2}}Alibaba Group\\
\IEEEauthorrefmark{0\text{3}}Mohamed bin Zayed University of Artificial Intelligence;    
\IEEEauthorrefmark{0\text{4}}Hamad Bin Khalifa University\\
jyangfan@u.nus.edu; \{feiwei, yaliang.li, bolin.ding\}@alibaba-inc.com\\ergute.bao@mbzuai.ac.ae; yyang@hbku.edu.qa; xkxiao@nus.edu.sg
}
}

\maketitle

\begin{abstract}
Given a table $T$ in a database and a question $Q$ expressed in natural language, the table question answering (TQA) task aims to return an accurate answer to $Q$ based on the content of $T$. 
The current state-of-the-art solutions leverage large language models (LLMs) to obtain high-quality answers. Most of these solutions, however, rely on proprietary, large-scale LLMs that require costly API access, which can be a significant financial barrier for many users. This paper focuses on TQA with smaller, open-weight LLMs that can run on a desktop or even a laptop. This is challenging since such LLMs typically have weaker capabilities compared to large proprietary ones, e.g., in terms of context understanding and instruction following. As a result, existing solutions suffer from a substantial performance drop when paired with such accessible LLMs.

We observe that one main reason for the poor performance of existing solutions with small open-weight LLMs is that these methods tend to ask the LLM to perform a highly sophisticated task with a long, complex prompt, which is often beyond the capabilities of such LLMs. Motivated by this, we present {\em \oursname}, a multi-agent approach designed to unlock the potential of accessible LLMs to enable high-quality, cost-effective TQA for a broader audience. The main idea of {\oursname} is to carefully coordinate a group of LLM agents, each performing a relatively simple task, through a structured, layered workflow to solve complicated TQA tasks --- akin to the coordination of an orchestra. This approach effectively reduces the complexity of prompts faced by each LLM agent, significantly improving the reliability of their outputs. 

We have implemented {\oursname} on top of AgentScope, an open-source multi-agent framework, and evaluated it across multiple TQA benchmarks using a wide range of open-weight LLMs. Experimental results demonstrate that {\oursname} achieves strong performance even with small- to medium-sized LLMs. For instance, when paired with Qwen2.5-14B, {\oursname} attains a test accuracy of 72.1\% on the WikiTQ benchmark, which approaches the best prior result 75.3\% achieved with GPT-4; meanwhile, when paired with a larger Qwen / Llama / DeepSeek model, {\oursname} beats all previous solutions and establishes new state-of-the-art results on all benchmarks in our experiments. The source code for {\oursname} is available at \url{https://github.com/Yangfan-Jiang/orchestra}.
\end{abstract}

\section{Introduction}\label{sec:intro}
Table question answering (TQA) aims to answer natural language queries based on tabular data, which presents an intuitive and intelligent interface for users to interact with a DBMS. TQA eliminates the need for the user to know the database schema or to manually write SQL queries, which is particularly convenient for applications such as financial report analysis, clinical decision support, and scientific research \cite{zhang2024reactable,zhu2024autotqa}. 
Given its significance, 
substantial research efforts \cite{li2014nalir,kim2020natural,ho2021qute,chengbinding,ye2023large,li2024table} have been devoted to developing effective TQA solutions. 
Yet, TQA remains a challenging task nowadays, due to the complexity of structured tabular data with an arbitrary schema, as well as the nuanced reasoning required to extract information, e.g., from textual columns.

Beyond its practical significance, TQA can be viewed as a natural extension of the classic problem of natural language interfaces to databases (NLIDB) \cite{hendrix1978developing,li2014nalir}, a long-standing area of interest in the database community. From this perspective, advancing TQA raises fundamental data management challenges, including robust query interpretation, handling of heterogeneous and semi-structured data, and ensuring responsible access and usage of sensitive information.

Recent advancements in large language models (LLMs) \cite{achiam2023gpt} have opened new opportunities for achieving effective TQA \cite{fernandez2023large}. Initially designed to process natural language inputs and generate coherent responses, LLMs have recently demonstrated remarkable reasoning capabilities \cite{wei2022chain,yaoreact,wang2023self}, leading to extensive research into their potential for tackling TQA tasks. In particular, recent studies \cite{wangtablechain,zhang2024reactable,zhu2024autotqa} have shown that enabling LLMs to interact with external tools, such as generating and executing SQL queries or Python code to extract relevant information from tables, yields state-of-the-art performance across various TQA benchmarks.

\subsection{Motivation} 
Existing LLM-powered TQA solutions typically rely on large, proprietary LLMs such as GPT-4 \cite{achiam2023gpt}, which involves 1.76 trillion parameters according to \cite{abacha2024medec}. Such LLMs usually exhibit strong capabilities for understanding long contexts, following complex instructions, as well as reasoning through chain-of-thought (CoT), which are pivotal to the success of the current state-of-the-art TQA algorithms. However, their use entails costly API calls, posing a significant barrier for small and medium-sized enterprises (SMEs) and individual users who regularly process tabular data but operate under tight budgets. Even when affordable, transmitting sensitive data to external APIs raises serious confidentiality and privacy concerns, particularly in high-stake applications such as analyzing internal financial statements or reviewing system logs for critical infrastructure.

{To address these issues, a promising alternative is to leverage more accessible LLMs, namely, open-weight models that are lightweight, locally deployable and efficient enough to run on modest hardware setups}, thereby eliminating both external API costs and associated privacy risks. The rapid advancements in open-weight LLMs, such as Qwen \cite{qwen2024qwen25technicalreport}, Llama \cite{dubey2024llama}, and DeepSeek \cite{liu2024deepseek}, alongside the development of efficient LLM inference and serving engines such as vLLM \cite{kwon2023efficient} and KTransformers \cite{ktransformers2024}, have made this approach increasingly accessible. In fact, it is now practical to serve large-scale open-weight LLMs on a single consumer-grade GPU \cite{ktransformers2024}. 

However, existing TQA solutions often fail to obtain high-quality results when paired with smaller, open-weight LLMs. 
In particular, to our knowledge, the state-of-the-art closed-source solution AutoTQA~\cite{zhu2024autotqa} only reports results with GPT-4; the best known open-source solution ReAcTable \cite{zhang2024reactable}, on the other hand, suffers substantial performance degradations when we replace the proprietary LLM with an open-weight one, as shown in Section \ref{subsec:exp-overall}. 
A key reason is that many existing methods employ long, complex prompts that exceed the capability of smaller LLMs, causing them to misinterpret instructions and produce incorrect or nonsensical outputs.

\subsection{Our Contributions}
In this paper, we present {\em \oursname}, a multi-agent framework for TQA, which remains highly effective with smaller, open-weight LLMs. 
Unlike prior frameworks such as ReAcTable~\cite{zhang2024reactable}, which rely on a single LLM instance and long, complex prompts, {\oursname} orchestrates multiple LLM agents, each handling a distinct, well-defined sub-task within a clear, concise context. These sub-tasks are designed to match the capabilities of current open-weight LLMs, ensuring that all agents contribute effectively to the final answer.
Notably, as we elaborate later in Section \ref{subsec:exp-overall}, several instantiations of {\oursname} with open-weight LLMs from the Qwen and Llama families achieve promising performance, matching or even surpassing the best prior results achieved by flagship proprietary models such as GPT-4.

To our knowledge, \oursname{} is the first work to formally study how small- to medium-sized open-weight LLMs can be orchestrated at inference time to solve TQA tasks effectively.
Our results reveal that, for TQA, allocating additional test-time computation to \emph{carefully orchestrated inference} is often more effective than simply increasing the backbone model size.
This observation extends recent findings on test-time scaling laws~\cite{snell2024scaling} beyond mathematical reasoning to TQA, a practically important yet comparatively underexplored setting.

\vspace{2mm}
\noindent\textbf{Challenges and key insights.}
The proposed {\oursname} framework focuses on tackling two fundamental limitations that arise when implementing the state-of-the-art TQA approaches with weaker LLMs. First, existing methods require the LLM to interact effectively with external tools such as SQL query engines and Python interpreters. Since many LLM lack native capabilities for tool use, such frameworks typically embed detailed instructions and few-shot demonstrations in the prompt. 
Second, these approaches typically involve multiple rounds of interaction with the same LLM, which further complicates the LLM's context. 
On the other hand, it is known that (i) the current generation of open-weight LLMs can be easily distracted by irrelevant context \cite{shi2023large}, and (ii) they struggle to follow intricate and multi-step instructions \cite{zengevaluating}. 
As a result, existing TQA frameworks using open-weight LLMs often lead to unreliable responses, with errors propagating through the multiple rounds of interactions.

Note that the above limitations may be intrinsic to small- to medium-sized open-weight LLMs since: (i) their attention mechanisms are less effective at filtering noisy context in long or dense inputs; (ii) their smaller parameter counts, according to the scaling law~\cite{kaplan2020scaling}, constrain their ability to capture complex patterns; and (iii) they generally lack extensive post-training processes, such as supervised fine-tuning (SFT) and reinforcement learning from human feedback (RLHF) \cite{ouyang2022training}, both of which are crucial for enhancing their instruction-following abilities and logical reasoning \cite{achiam2023gpt}. 
Consequently, directly applying existing reasoning techniques, such as chain-of-thought (CoT) \cite{wei2022chain} and ReAct \cite{yaoreact}, may be insufficient for effective TQA with these models.

In this paper, we conduct an in-depth analysis of the workflow in state-of-the-art TQA frameworks \cite{wangtablechain,zhang2024reactable}, which reveals that the complicated reasoning process in these frameworks essentially consists of a pair of simpler and largely independent reasoning paths. Specifically, one path focuses on logical derivations, progressively extracting and refining table information toward answering the query, while the other path determines how to interact with external tools to extract relevant information from the database. These insights suggest that the complex reasoning flow of TQA can be decomposed into smaller subtasks, and assigned to multiple compartmentalized LLM-based agents, which is the main intuition behind the proposed approach {\oursname}.

\vspace{2mm}
\noindent\textbf{Solution overview.} 
Based on the above insights, we first present a two-agent solution that splits the TQA process into two simpler subtasks: (i) a {\emph{logic agent}} that handles logical derivations based on the user's TQA query using data extracted from the table, and (ii) a {\emph{query agent}} that manages analytical operations, generating SQL queries and/or Python code to interact with external tools, and precisely extract the required information from the table following the logic agent's instructions. Each agent is instantiated with concise, role-specific prompts tailored to its subtask.

While the logic and query agents follow distinct reasoning paths, they must interact strategically to maintain consistency and avoid error propagation, e.g., due to hallucinations \cite{ji2023survey}. Specifically, the logic agent requires the supporting evidence extracted by the query agent to make grounded decisions; and the latter, in turn, needs clear directions from the former. 
To achieve this, {\oursname} employs in-context learning \cite{brown2020language} to orchestrate the interaction, synchronizing the agents' states at each step to ensure coherent progress toward an accurate and reliable answer. 

As we elaborate further in Section \ref{sec:our-solution}, the above two-agent system  
still faces two challenges: 
(i) irrelevant information remains in the reasoning context, typically embedded in the few-shot prompts used to initiate in-context learning, which may confuse the agent; (ii) the formalization of the two-agent system's overall objective may deviate from the original TQA task goal over time.

The proposed solution {\oursname} is built on the above two-agent approach and addresses these two issues through careful algorithmic designs, as follows.
First, {\oursname} refines the reasoning context by filtering out irrelevant information such as few-shot prompts, ensuring that only TQA task-relevant details remain. It also introduces {a third LLM instance}, called the {\emph{decision agent}}, which directly derives the final answer based on this carefully refined reasoning context. To tackle the second issue, we formally analyze the misalignment between the objective of the above first-cut solution and the original TQA task goal. Our analysis shows that this misalignment can be effectively calibrated using a Monte Carlo method, ensuring closer adherence to the original TQA task.

\vspace{2mm}
\noindent\textbf{Implementation and evaluations.} We implement {\oursname} using AgentScope \cite{gao2024agentscope}, an open-source multi-agent framework, and conduct extensive evaluations across three well-adopted TQA benchmarks: WikiTQ \cite{pasupat2015compositional}, TabFact \cite{chentabfact}, and TableBench \cite{wu2025tablebench}. Our evaluation covers a wide range of open-weight LLMs, including models from Qwen \cite{qwen2024qwen25technicalreport}, Llama \cite{dubey2024llama}, DeepSeek \cite{liu2024deepseek}, Gemma \cite{team2024gemma}, and Mistral \cite{jiang2023mistral}. 
The results demonstrate that {\oursname} consistently and significantly enhances the performance of open-weight LLMs on TQA tasks.
For example, when instantiated with the Qwen 14B model, {\oursname} achieves a test accuracy of 72.1\% on the WikiTQ benchmark, outperforming the leading TQA framework based on the same model by a substantial margin of 11.4\%. 
Moreover, our evaluation shows that {\oursname}, when paired with larger open-weight LLMs such as the 70-billion-parameter models from the Qwen and Llama families, surpasses the best prior results achieved by GPT-4 in almost all cases, establishing new state-of-the-art results.

\section{Preliminaries}
\label{sec:preliminaries}
\subsection{Table Question Answering}
Given a table $T$ and a natural language question $Q$, table question answering (TQA) seeks to produce an accurate and reliable answer to $Q$. TQA finds important applications in various fields, such as financial report analysis, clinical decision support, and scientific research \cite{zhang2024reactable,zhu2024autotqa,zhang2024finsql}. Designing a general-purpose, end-to-end solution for TQA is challenging due to the diverse structures and semantics of tables, the wide range of content (e.g., in specialized domains such as finance to medicine), and the varying logical reasoning and numerical analysis required. %

\begin{figure}[!t]
\centering
\hspace{-0mm}\includegraphics[width=1.0\linewidth]{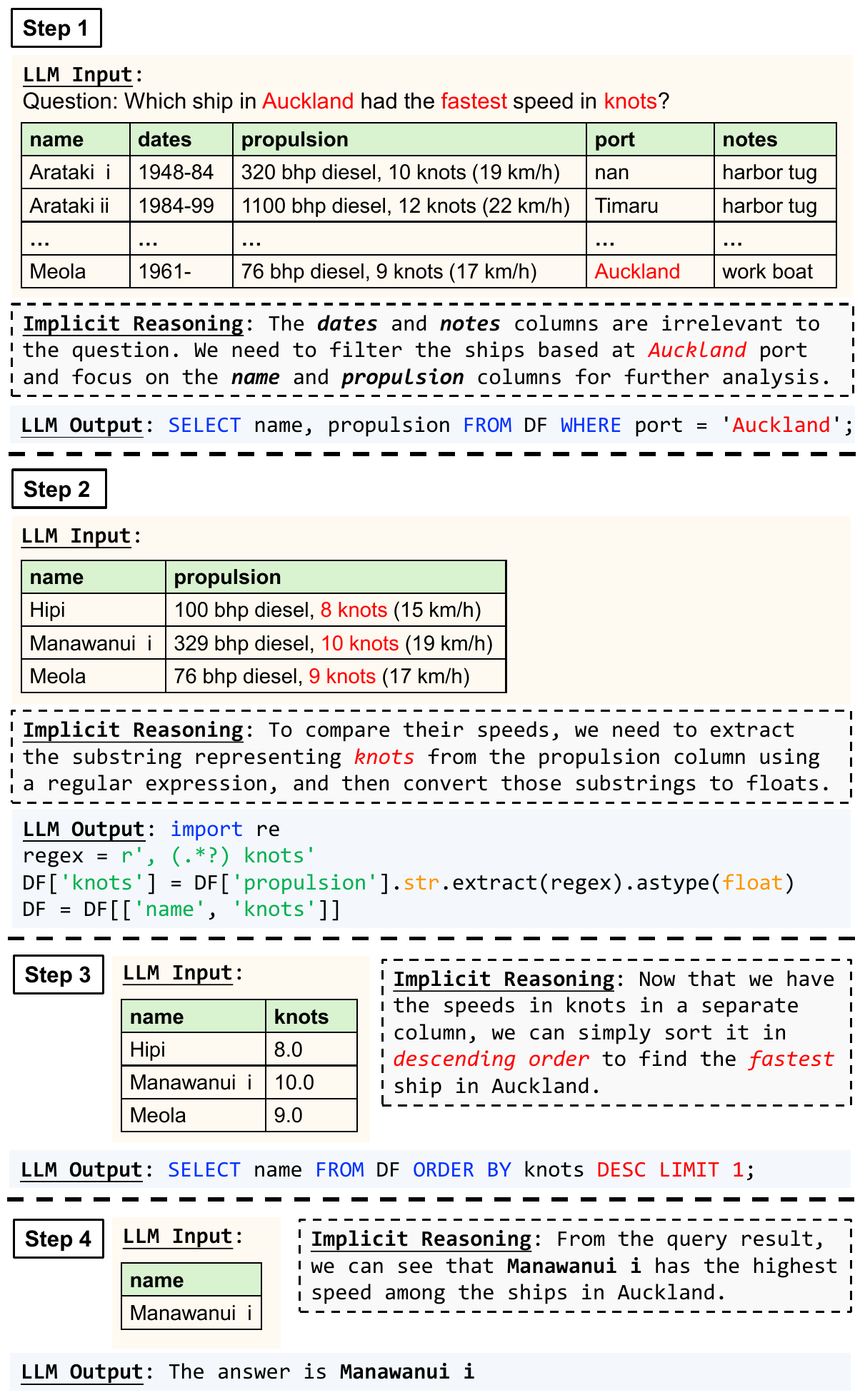}
\vspace{-2mm}
\caption{{Running example of answering a TQA query through multiple rounds of LLM reasoning, database interactions via SQL, and postprocessing with Python. Few-shot prompts used for in-context learning (see Section \ref{subsec:cot_react} for details) are omitted for clarity.}
}
\label{fig:sota-work-flow}
\vspace{-1mm}
\end{figure}

Figure~\ref{fig:sota-work-flow} illustrates an example TQA: given a table listing various attributes of ships, the task is to determine the name of the fastest Auckland-based ship. One possible workflow, shown in the figure, involves the following steps (i) filtering ships at Auckland, (ii) extracting the speed for each such ship, (iii) sorting the ships by speed and (iv) returning the fastest ship. In this example, the TQA
engine needs to understand that ``Auckland'' is a value of the \textit{port} attribute, and the speed of the ship can be obtained from the \textit{propulsion} column through string manipulations. Note that the TQA engine has no prior information about the table schema, and, thus, must derive the above from the table records. Meanwhile, 
it needs to compose correct SQL queries to retrieve relevant data from the DBMS, as well as a Python script to extract speed information from the retrieved data.

\subsection{Large Language Models for TQA}\label{subsec:llm}
Large language models (LLMs) are deep neural networks trained on vast amounts of text data, enabling them to understand and generate texts with impressive proficiency. They are typically built on the Transformer architecture, which utilizes attention mechanisms \cite{vaswani2017attention} to capture linguistic patterns, structures, and semantics.  
LLMs typically consist of billions or even trillions of parameters \cite{qwen2024qwen25technicalreport,dubey2024llama,liu2024deepseek}, and a higher parameter count often leads to stronger capability to handle diverse and complex tasks, according to the LLM scaling law~\cite{kaplan2020scaling}. %
Recent research has demonstrated the strong potential of LLMs in tackling data analysis tasks \cite{fernandez2023large,zhang2024finsql,luoma2025snails,giannakouris2025lambda,yan2024gidcl,li2025llm,ren2024purple,fan2024metasql,zheng2024adapting,zhang2025clear,fan2025grounding,li2025aidsql}. 

Early solutions for TQA with LLMs are often based on the methodology of directly translating the natural language query in TQA into executable code such as SQL or Python \cite{wangtablechain,li2024table,gao2024text,trummer2024generating,fan2024combining}, which extracts relevant information from the table to form the final answer. 
The main limitation of this methodology is that it aims to generate code to answer the entire TQA query \textit{in one shot}. 
For complex TQA queries, the appropriate code often depends on the data, meaning that multiple interactions with the database are often needed. 
In our running example shown in Figure \ref{fig:sota-work-flow}, intuitively, one needs to read some example data records to understand that ship speed can be extracted from the ``propulsion'' attribute (e.g., ``320 bhp diesel, 10 knots (19 km/h)''), using a regular expression ``\texttt{,(.*?) knots}''. 
In other words, dynamic interactions with the table data, as well as multi-step reasoning, are needed to answer this TQA query, which are beyond the capabilities of a static, single-shot text-to-code translator. 
Consequently, such methods often lead to sub-optimal performance in TQA tasks, as shown in \cite{chengbinding,li2024table,wangtablechain}. 
 
To address these challenges and achieve effective TQA, a promising approach is to leverage the advanced reasoning capabilities of LLMs, described in the next subsection.

\subsection{Reasoning-Enhancing Techniques for LLMs}\label{subsec:cot_react}
\noindent\textbf{Chain-of-Thought (CoT) \cite{wei2022chain}.} 
CoT is a general approach for enhancing the reasoning abilities of LLMs. The main idea is to prompt LLMs to conduct the reasoning process by generating a reasoning context before arriving at the final answer. Specifically, in the CoT paradigm, LLMs are guided to produce intermediate reasoning context step by step, attempting to solve a question progressively. This approach is particularly effective for tackling complex tasks that inherently require multiple reasoning steps to derive the correct answer, such as solving mathematical problems, logical puzzles, or table-related questions \cite{achiam2023gpt,wang2023self,zhang2024reactable,qwen2024qwen25technicalreport,zhang2025survey}.

\vspace{2mm}
\noindent\textbf{Reasoning and acting (ReAct) \cite{yaoreact}.} 
ReAct is a paradigm that synergizes reasoning and acting within LLMs. It augments the reasoning process by allowing LLMs to interact with external tools and environments during problem-solving. Since LLMs are essentially probabilistic language models, they are not explicitly trained to perform numerical computations or operations on tabular data. 
For instance, previous studies have shown that even state-of-the-art LLMs struggle with basic tasks such as identifying the number of rows or columns in a table or comparing numerical values \cite{li2024table}. 
ReAct addresses this limitation by explicitly prompting the LLM to interact with external tools, thereby supplementing its internal reasoning with precise computations and data handling. This interaction can be viewed as an extension of the CoT, where each reasoning step is augmented with feedback from external functions. Notably, state-of-the-art TQA solutions \cite{wangtablechain,zhang2024reactable} have employed the ReAct paradigm to obtain high-quality TQA results. %

\vspace{2mm}
\noindent\textbf{Few-shot learning and reinforcement learning.}
A common method for initiating reasoning-enhancing techniques is through in-context learning with few-shot prompts. In this approach, LLMs are provided with carefully crafted examples that include the reasoning process leading to the correct answer. The LLM then imitates this reasoning process when presented with new inputs. This strategy has been commonly adopted in existing TQA solutions.
{
For example, one may manually construct multiple examples similar to those in Figure~\ref{fig:sota-work-flow}, prompt the LLM with these examples, and have it imitate the TQA process shown in the figure, i.e., processing the table step-by-step by invoking tools to solve TQA.
}

Finally, reinforcement learning (RL) has been shown as an effective post-training method for LLMs to automatically optimize reasoning trajectories \cite{schulman2017proximal,ouyang2022training,rafailov2023direct}. For instance, OpenAI o1 \cite{openaio1}, DeepSeek-R1 \cite{guo2025deepseek}, and QwQ \cite{qwq2025} use RL to align intermediate reasoning steps with task objectives, rewarding coherent logic and correct conclusions.
Applying RL to TQA, however, is rather challenging, since the TQA reasoning chain typically involves iterative interactions with the database system to retrieve data (e.g., in our running example in Figure \ref{fig:sota-work-flow}, the LLM needs to retrieve example attribute values to determine the regular expression used for extracting speed information), unlike other reasoning tasks such as math or coding. This may complicate the design of the RL reward function. Further RL training for LLMs necessitates substantial costs in terms of computing power, especially for larger models. Thus, we leave RL for TQA as future work.

\section{{\oursname}}\label{sec:our-solution}
As explained in Section \ref{sec:intro}, the key insight behind the design of {\oursname} is that the complex TQA process in the current state-of-the-art methods can be decomposed into two simpler subtasks: logical derivations and data processing. The former involves the LLM performing logical reasoning based on evidence extracted from the table and determining what additional steps are needed to arrive at the final answer. The latter focuses on processing data from the database, e.g., through SQL queries and/or Python scripts. 

In a nutshell, the proposed solution {\oursname} sets up a \emph{logic agent} and a \emph{query agent} for these two subtasks, respectively. 
The two agents work interactively, adaptively updating their states as they progress through their respective subtasks. Once the logic agent reaches a stage where the final answer is imminent, {\oursname} halts the two-agent system and obtains the context composed by the logic agent, in order to infer the final answer. 
Subsequently, {\oursname} refines this context by removing irrelevant information, ensuring its conciseness and clarity. 
To do so, a \emph{decision agent} is introduced to generate the TQA answer based on the refined reasoning context. Finally, a Monte Carlo method is applied to further calibrate the output of the decision agent, ensuring that the final answer aligns with the original goal of the TQA task.

In the following, Section \ref{subsec:two-reasoning-paths} elaborates on the rationale behind the decomposition of the TQA workflow. Section \ref{subsec:two-agent} focuses on the design of the logic and query agents, as well as their interactions.
Section \ref{subsec:calibrate-obj} presents effective optimizations to improve the final TQA answer. Section \ref{subsec:discussions} discusses several important design decisions in {\oursname}.

\subsection{Decomposing the TQA Workflow}\label{subsec:two-reasoning-paths}

We begin by examining the workflow of LLMs in the current state-of-the-art TQA approaches \cite{wangtablechain,zhang2024reactable,zhu2024autotqa}, which combine CoT and ReAct (explained in Section \ref{subsec:cot_react}) to iteratively reason over and interact with the tabular data. Figure \ref{fig:sota-work-flow} illustrates a typical workflow of these methods for our running example. 
The process starts by prompting the LLM with sample records of the given table and the corresponding TQA question in natural language. In each step, the LLM is expected to perform actions that extract relevant insights from the table. For instance, in Step 1 in Figure \ref{fig:sota-work-flow}, the LLM's input context contains
the question and the records as shown in the figure; then, the LLM is expected to observe the records, and to (implicitly) reason similarly as the following:

``\texttt{\small The dates and notes columns are irrelevant to the question. We need to filter the ships based at \textbf{Auckland} port and focus on the \textbf{name} and \textbf{propulsion} columns for further analysis}.''

With the above goal in mind (i.e., filtering the name and propulsion columns for ships based at Auckland port), the LLM is then expected to take action by composing the following SQL query, and submitting it to the DBMS:
\vspace{-2pt}
\lstset{upquote=true}
\lstdefinestyle{mystyle}{
    commentstyle=\color{green},
    keywordstyle=\color{blue},
    stringstyle=\color{purple},
    basicstyle=\small\ttfamily,
    breaklines=true,
    columns=fullflexible,
    frame=single,
}
\lstset{style=mystyle}
\begin{lstlisting}[language=SQL,
	deletekeywords={IDENTITY},
	deletekeywords={[2]INT},
	morekeywords={clustered},
	mathescape=true,
	xleftmargin=0pt,
	framexleftmargin=0pt,
	frame=tb,
	framerule=0pt]
   SELECT name, propulsion FROM DF
   WHERE port='Auckland';
\end{lstlisting}
\vspace{-2pt}
The return values of the above SQL code are then used to prompt the LLM in the next iteration. 

Observe that in the above step, there is a clear \textit{context switch}: the LLM's initial input context includes table records and users' queries, and its output contains an SQL statement; in other words, the LLM switches from understanding records to writing SQL. Further, in later steps of the workflow, both of the above (i.e., data records, SQL code) are included in the LLM's input context, in order for the LLM to understand its past interactions with the database. This reveals a major drawback of existing LLM-based TQA frameworks: that \textit{they do not maintain a logically coherent reasoning context}. In particular, the reasoning chain is fragmented, as the LLM must frequently switch between different tasks, e.g., observing the retrieved data, deciding on what to do next, writing/running/debugging code, etc. Consequently, the LLM prompt is rather long and complex, with frequent context switching, which is difficult to process correctly, especially for smaller, open-weight LLMs.

Continuing the running example, in the next step, the LLM's input also includes the data records, SQL query, and its results from Step 1. The LLM is expected to output Python code that extracts speed information using regular expressions, which will be included in the LLM's context in future steps. In general, the contents of the LLM's context alternates between retrieved data and generated code, which tends to confuse the LLM, especially weaker ones with a smaller parameter count.

Meanwhile, the above example reveals that in the TQA process, there are two interleaved reasoning paths, each serving a distinct purpose: (i) deciding on what to do next based on the obtained data and previous actions, and (ii) writing code to interact with the database and post-process the retrieved data. Note that the latter reasoning path does not focus on the TQA task per se; rather, it aims to generate semantically correct code that accurately represents the actions required by the first path. Since existing TQA solutions do not distinguish these two fundamentally different reasoning processes, the LLM suffers from frequent context switching as well as fragmented, confusing inputs, which adversely affects the clarity and coherence of its reasoning flow.
These pose significant challenges for smaller open-weight LLMs, which typically have limited reasoning and context comprehension capacities, as reviewed in Section \ref{subsec:llm}.

To address this issue, we propose decomposing the TQA reasoning process into two subtasks, allocated to specialized LLM agents, with each agent dedicated to one of the reasoning paths. The agent designs and their interactions are clarified in the next subsection.

\begin{figure*}[!t]
\centering
\hspace{-0mm}\includegraphics[width=1.0\linewidth]{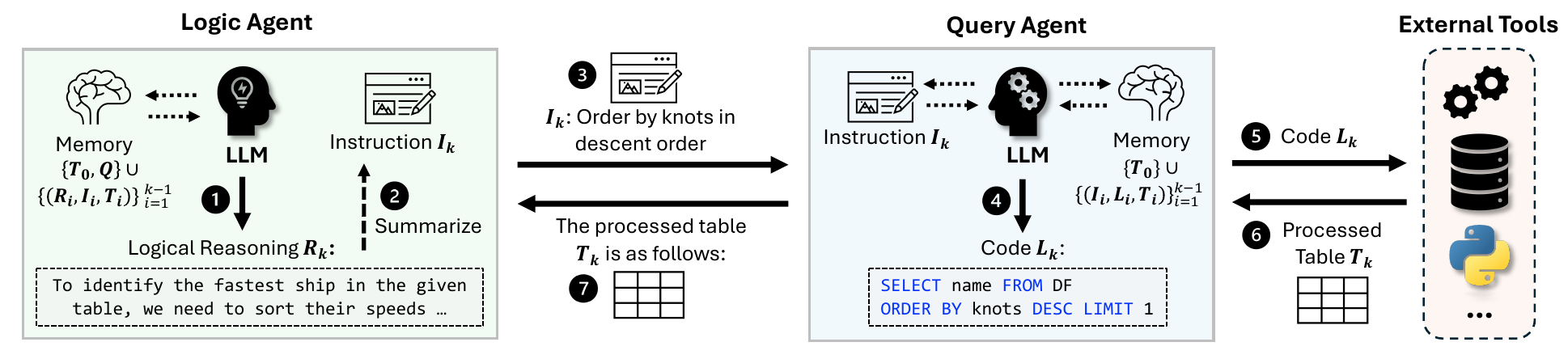}
\vspace{-3mm}
\caption{Example interactions of the logic and query agents.}
\label{fig:two-agent}
\vspace{-1mm}
\end{figure*}

\subsection{Logic and Query Agents}\label{subsec:two-agent}
Since the TQA workflow involves two interleaved reasoning paths, for logic derivations and data querying, respectively, we first present a two-agent system consisting of a \emph{logic agent} and a \emph{query agent}. To effectively compartmentalize agents, we restructure these two reasoning paths to ensure that each one is self-contained. 
The key to achieving this is to \textit{make the implicit reasoning explicit} in each step of the workflow. For instance, in our running example in Figure \ref{fig:sota-work-flow}, the reasonings shown in dashed boxes are now materialized by the logic agent, and translated into clear, actionable instructions for the query agent to compose SQL/Python code.

Figure \ref{fig:two-agent} illustrates an interaction round $k$ in this two-agent system. Specifically, the logic agent reviews the conversation history stored in its memory and determines what information is needed from the table to facilitate further logical derivations for answering the TQA query. It then generates an intermediate reasoning statement $R_k$, specifying the supporting data needed and the actions required to obtain such data. Subsequently, a clear and concise instruction $I_k$ is summarized from $R_k$ and sent to the query agent.

Upon receiving the instruction $I_k$, the query agent translates the specified actions in $I_k$ into executable code $L_k$, e.g., SQL queries and/or Python scripts. The code $L_k$ is then executed with external tools, which return an intermediate table $T_k$ containing the supporting evidence required by $R_k$. $T_k$ is subsequently sent back and accessed by both the logic and query agents. 

At the end of round $k$, each agent updates its respective memory with the newly observed and generated information (i.e., $R_k$, $I_k$, and $T_k$ for the logic agent; $I_k$, $L_k$, and $T_k$ for the query agent) and proceeds to the next round $k+1$. Once the logic agent has accumulated sufficient evidence and is confident in the final answer to the TQA query, it outputs that answer and terminates the interaction process. 
Below, we clarify the detailed designs of the two agents.

\vspace{2mm}
\noindent\textbf{Logic agent.} The logic agent is composed of an LLM and a memory module. The memory is initialized with the original TQA input, which includes table $T_0$ and the corresponding question $Q$. The LLM is prompted to reason toward answering the question $Q$ and interact with the query agent to retrieve supporting evidence from the table as needed. In each iteration $k$, the logic agent either chooses to interact with the query agent by issuing an instruction $I_k$ or produces the final answer to the TQA task if the supporting evidence in its memory provides sufficient confidence in the conclusion. Unlike LLMs in prior TQA frameworks, the LLM in the logic agent only observes the most relevant information for the TQA task. It does not access irrelevant data, such as code snippets or interaction logs with external tools, thus maintaining a focused reasoning context and enhancing the reliability of its results at each step.

\vspace{2mm}
\noindent\textbf{Query agent.} The query agent serves as an interface that executes actions specified by the logic agent's instructions and returns the results to support further reasoning. Similar to the logic agent, the query agent comprises an LLM and a memory module to maintain context throughout the workflow. The key distinction lies in its integration with external tools, e.g., DBMS and/or Python interpreter. Upon receiving an instruction $I_k$, the query agent first reviews its memory, which stores both the interaction history and the table to be processed. It then translates $I_k$ into code $L_k$ and invokes the corresponding external tool to run $L_k$. After obtaining the result (often in non-text formats, such as DBMS query outputs or Pandas DataFrames) the query agent converts it into a text-based table $T_k$ (e.g., in Markdown format) compatible with LLMs, which is then shared with the logic agent, facilitating further reasoning.

\vspace{2mm}
\noindent\textbf{Calibrating the behavior pattern.} 
To ensure that the LLMs in both the logic and query agents adhere to the intended interaction design, we employ in-context learning techniques with carefully crafted few-shot prompts to calibrate their behavior. This approach keeps each LLM focused on its designated role, preventing irrelevant or unexpected actions while enhancing reliability. Additionally, in-context learning effectively control the granularity of each reasoning step, ensuring that sub-tasks in each interaction round remain manageable and within the capabilities of the LLMs.

Since {\oursname} simplifies the process by assigning each LLM agent a well-defined and straightforward sub-task, it is able to use more comprehensible few-shot prompts for in-context learning, while remaining within the capabilities of small- to medium-sized open-weight LLMs. Next, we further improve upon this two-agent system with optimizations for refining the final answer.

\subsection{Refining the Final Answer}\label{subsec:calibrate-obj}
The two-agent system outlined above serves as our first-cut solution to tackling TQA tasks by decomposing the complex TQA workflow into subtasks that fit within the reasoning and instruction-following capabilities of open-weight LLMs. 
This setup, however, also introduces two new issues, elaborated below, that may lead to a misalignment between the system's output and the original TQA query. As a result, directly adopting the answer produced by the logic agent may yield suboptimal performance. To address these issues, we further optimize our TQA framework with strategies that effectively realign the TQA framework with its intended objective.

\vspace{2mm}
\noindent\textbf{Refining the final output with a decision agent.}
The first issue arises from the complex few-shot prompts used to guide the logic agent. While these prompts help enforce step-by-step reasoning, they consume a large portion of the input context and may distract the model without contributing directly to the TQA task.  
An ideal solution would be to guide the LLM to reason step-by-step within only the most relevant context, ensuring that the final answer is conditioned solely on the actual reasoning process. However, this creates a conflict: on one hand, few-shot prompts are necessary to facilitate in-context learning to shape the logic agent's behavior; on the other hand, we aim for the logic agent to produce the final answer based solely on the reasoning context, excluding the unnecessary few-shot prompts. 

To resolve this conflict, we introduce a third LLM agent called the \emph{decision agent} in the {\oursname} framework. This agent ensures that the final answer is conditioned exclusively upon the most relevant reasoning context, without being influenced by irrelevant few-shot prompts or other instructions.  
The process works as follows: first, the logic and query agents interacts until the logic agent produces a preliminary answer.  
Next, we extract its reasoning trace, filtering out irrelevant elements such as few-shot examples and instructions, and use the refined context to initialize the decision agent's memory. The decision agent then produces the final answer based solely on this focused reasoning context.

\vspace{2mm}
\noindent\textbf{Calibrating the final objective.}
The second issue reflects a more fundamental misalignment between the two-agent framework and the original TQA objective.
Formally, the goal of TQA is to determine the answer $A$ that maximizes the likelihood given a table $T_0$ and a question $Q$:
\begin{align}
    \text{Answer}=\arg\max_{a} \Pr\left[A=a \mid T_0, Q \right].\label{eq:tqa-initial-goal}
\end{align}
However, the {\oursname} framework described so far first samples an intermediate reasoning context through agent interactions, and the decision agent then produces the final answer based on this context, leading to the following formulation: 
\begin{align*}
    \text{Answer}=\arg\max_{a}\Pr\left[A=a \mid r^*, T_0, Q\right] \cdot \Pr\left[R=r^*\mid T_0, Q\right],
\end{align*}
where $r^*=\arg\max_{r} \Pr\left[R=r \mid T_0, Q\right]$ represents the intermediate reasoning context generated by the logic agent.

This formulation deviates from Eq.~(\ref{eq:tqa-initial-goal}) by introducing a dependency on a single selected reasoning path. To rectify this misalignment, a more principled approach is to marginalize over all possible paths:

\begin{align}
   & \arg\max_a  \Pr\left[ A=a \mid T_0,Q \right] \nonumber \\
   = & \arg\max_a  \sum_r \underbrace{\Pr\left[A=a \mid R=r,T_0,Q \right]}_{\text{decision agent}} \cdot \underbrace{\Pr\left[ R=r \mid T_0,Q \right]}_{\text{two-agent system}}.\label{eq:sum-over-all-paths}
\end{align}
In this formulation, the term within the summation reflects our workflow: the two-agent system samples a reasoning path $R=r$ given $T_0$ and $Q$, and the decision agent generates the final answer conditioned on $r$. Aligning with the original TQA objective therefore requires marginalizing over all possible reasoning paths. 
Since enumerating all paths is computationally infeasible, we instead approximate the objective using Monte Carlo methods. By interpreting the summation as an expectation, we reformulate Eq.~(\ref{eq:sum-over-all-paths}) as:
\begin{align}
    &\arg\max_a \sum_r \Pr\left[A=a \mid R=r,T_0,Q \right] \cdot \Pr\left[ R=r \mid T_0,Q \right] \nonumber\\
    =&\arg\max_a \mathbb{E}_{r\sim \Pr[R\mid T_0, Q]}\big[\Pr\left[A=a\mid R=r, T_0, Q\right]\big] \label{eq:sum-2-expect}.
\end{align}
This allows us to approximate the summation by estimating the expectation of $\Pr[A=a\mid R=r, T_0, Q]$ over $r$, which can be efficiently done with Monte Carlo sampling. Specifically, we first sample $m$ reasoning paths from the distribution $\Pr[R\mid T_0,Q]$ and average their answer probabilities, resulting in the following approximate formulation of the realigned objective:
\begin{align}
    \arg\max_a \frac{1}{m} \sum_{i=1}^{m} \Pr\left[A=a \mid r_i,T_0,Q \right],\label{eq:mc-estimation}
\end{align}
where each $r_i$ is independently sampled from $\Pr[R \mid T_0, Q]$. 
{In practice, we approximate Eq.~(\ref{eq:mc-estimation}) through Monte Carlo sampling: the two-agent system is executed $m$ times to sample reasoning paths, each evaluated by the decision agent, and the dominant answer is selected as the final output.}

\vspace{2mm}
\noindent\textbf{Putting it all together.} By combining these components, we establish the complete {\oursname} framework, which begins by running the two-agent system described in Section \ref{subsec:two-agent} independently $m$ times to sample $m$ reasoning paths $\{r_1,\dots,r_m\}$. For each reasoning path $r_i$, a decision agent is instantiated accordingly to generate a corresponding candidate answer $a_i$, conditioned on $r_i$, the table $T_0$, and the question $Q$. Finally, the most frequently occurring answer in the candidate set $\{a_1, \dots, a_m\}$ is selected as the final output, ensuring alignment with the original task objective.

\subsection{Discussions}\label{subsec:discussions}
In the literature, several techniques (detailed below) are commonly applied to enhance LLMs' ability to understand tabular data in previous work \cite{li2024table,zhu2024autotqa,wang2023self}, which are absent in the proposed solution {\oursname}. Below, we briefly discuss why  {\oursname} excludes them.

\vspace{2mm}
\noindent\textbf{Permutation invariance.}
Permutation-invariant properties have been leveraged in data augmentation to train 
LLMs specifically for tabular data \cite{li2024table}. The intuition is that permuting the rows or columns of a table should not alter its semantic meaning. Accordingly, one could randomly shuffle table rows and columns, have {\oursname} generate answers for each variation, and select the most frequently occurring answer as the final output. However, we note that this method may not be suitable for all TQA tasks. In particular, table rows are often ordered according to a specific criterion, such as descending years, while columns may contain aggregate information, e.g., the last column may summarize values from other columns. In such cases, random permutations could disrupt meaningful structures, leading to incorrect reasoning. For this reason, we do not incorporate permutation-invariant techniques into {\oursname} by default, to ensure that {\oursname} achieves reliable performance out-of-box. 
Nonetheless, our modular design allows advanced users to incorporate such techniques for specific use cases.

\vspace{2mm}
\noindent\textbf{Verifying final answer.} 
Another common technique for improving LLM performance, particularly in reasoning tasks involving mathematics or numerical problems, is verifying the final answer using an additional LLM agent as a verifier~\cite{kambhampatiposition,zhu2024autotqa}. However, we argue that this method is not well-suited for TQA in general. Answer verification is particularly effective when the problem-solving process is asymmetric, where verifying an answer is significantly easier than computing it. For example, solving a polynomial equation may be complex, but checking its correctness through substitution is straightforward. In contrast, TQA tasks lack this asymmetry. The process of verifying an answer in TQA is often as complex as deriving it in the first place, meaning a verification agent would struggle to assess correctness without essentially solving the TQA problem itself. Therefore, we do not incorporate a verification agent into the {\oursname} framework.

\vspace{2mm}
\noindent\textbf{Voting.} 
Voting mechanisms have been used in ReAcTable, a previous LLM-based TQA framework \cite{zhang2024reactable}, to address ambiguity and improve answer quality. Specifically, ReAcTable explores three voting mechanisms: majority voting, tree-exploration voting, and execution-based voting. We note that the implementation of majority voting is similar to our Monte Carlo method. However, ReAcTable does not provide strong insights or analysis regarding the rationale behind these mechanisms. Additionally, its workflow fundamentally differs from that of {\oursname}, meaning that directly incorporating these voting mechanisms into {\oursname} may not yield an effective solution. In contrast, through formal analysis, we demonstrate the feasibility of our Monte Carlo sampling process. Furthermore, we observe that tree-exploration or execution-based voting do not align well with the original TQA objective expressed in Eq. (\ref{eq:tqa-initial-goal}). This misalignment may partially explain why prior evaluations \cite{zhang2024reactable} have shown that these two voting mechanisms fail to improve TQA performance.

\section{Experiments}\label{sec:experiments}
This section empirically evaluates the performance of {\oursname} on three widely adopted TQA benchmarks. Our evaluations demonstrate that when paired with the same  open-weight LLMs, {\oursname} consistently and significantly outperforms ReAcTable \cite{zhang2024reactable}, the state-of-the-art LLM-powered TQA framework. Moreover, when instantiated with large open-weight models such as Qwen2.5-72B, Llama3.1-70B, and DeepSeek-V3, {\oursname} surpasses prior best results achieved with GPT-4, establishing a new state of the art. 

\subsection{Setup}\label{subsec:exp-setup}
We have implemented the {\oursname} framework using AgentScope \cite{gao2024agentscope}, an open-source multi-agent framework. The detailed experimental setup is described below.

\vspace{2mm}
\noindent\textbf{Benchmark.} We use three well-established TQA benchmarks in our experimental evaluations: WikiTQ \cite{pasupat2015compositional}, TabFact \cite{chentabfact}, and TableBench \cite{wu2025tablebench}.

\begin{itemize}[leftmargin=*]
    \item {\bf WikiTQ} is a representative benchmark for general TQA tasks, containing 4,344 table-question-answer triples in the test set. 
    It includes complex questions requiring multiple reasoning steps and various data operations, such as comparison, aggregation, and numerical computation. The ground-truth answers are typically concise and objective.
    \item {\bf TabFact} is a standard table-based fact verification benchmark. It contains 1,998 test cases, each associated with a table and manually annotated statements related to it. The answers are binary, i.e., either ``yes'' or ``no'', indicating whether the given statement is true or false based on the tabular data.
    {\item {\bf TableBench} is a recently released benchmark that is comprehensive and substantially more complex than earlier datasets. It covers a wide range of domains, including finance, infrastructure, education, transportation, and healthcare, making it more representative of real-world and industrial scenarios than prior academic datasets.} 
\end{itemize}

\vspace{2mm}
\noindent\textbf{Metrics.}
We use evaluation metrics identical to those in prior LLM-powered TQA frameworks \cite{wangtablechain,zhang2024reactable,zhu2024autotqa}. 
Specifically, we use accuracy to measure the percentage of generated outputs that exactly match the ground truth answers. 

\vspace{2mm}
\noindent\textbf{Baselines.}
We compare {\oursname} with ReAcTable \cite{zhang2024reactable}, the state-of-the-art LLM-powered TQA framework. We use the code provided in \cite{zhang2024reactable} and keep all settings as default. Additionally, we include a chain-of-thoughts (CoT) \cite{wei2022chain} baseline, which prompts a single LLM to solve TQA by reasoning step by step without invoking external tools. We also note that a concurrent work, Chain-of-Table, \cite{wangtablechain}, employs LLMs for TQA tasks. However, it shares a similar algorithmic design as ReAcTable, driving a single LLM to interact with external tools step by step for table processing. ReAcTable enables the LLM to generate arbitrary SQL queries and Python scripts, whereas Chain-of-Table restricts the LLM to a predefined set of SQL operations. Due to its expanded capabilities, ReAcTable is reported to outperform Chain-of-Table across all three TQA benchmarks \cite{zhang2024reactable,zhang2025survey}. Therefore, in our experiments, we focus on comparing {\oursname} against ReAcTable. 

{Note that to ensure a fair and competitive comparison, we further optimized the ReAcTable system prompt by (i) requiring ``SQL:'' or ``Python:'' prefixes for code, (ii) enclosing code in ``\texttt{```}'' delimiters, and (iii) formatting final answers to match task‑specific constraints (e.g., concise factual answers for WikiTQ and exactly ``yes'' or ``no'' for TabFact). We evaluated both the original ReAcTable and our optimized version, and found that adding the tuned system prompt consistently improved TQA performance. All subsequent ``ReAcTable'' results reported in this paper refer to this optimized version, which more accurately reflects the framework's capabilities.}

Additionally, a recent work AutoTQA \cite{zhu2024autotqa} proposes an LLM-based TQA framework. {However, AutoTQA is primarily designed for multi-table, domain-specific scenarios, while Orchestra focuses on single-table, general-purpose TQA. Moreover, AutoTQA is a proprietary industrial product, and its implementation is not publicly available.} Consequently, we are unable to conduct a direct comparison with AutoTQA under the same LLM. Instead, we mention the reported AutoTQA result \cite{zhu2024autotqa} obtained using GPT-4 whenever applicable.

\vspace{2mm}
\noindent\textbf{Configurations.} We evaluate the TQA performance of {\oursname} using a wide range of LLMs. %
To realign {\oursname}'s TQA objective using Monte Carlo sampling, we set the temperatures of all LLMs to $0.7$ across all experiments, ensuring diversity in reasoning paths across different trials. By default, we set the number of Monte Carlo simulations to $5$ for all benchmarks.

{\oursname} requires the logic agent and query agent to interact until the logic agent generates an answer. To prevent excessively long TQA processes or infinite loops caused by unexpected outputs, we cap the maximum number of interaction rounds at $5$. Once the number of interactions reaches $5$, we prompt the logic agent with ``Please provide an answer directly'', forcing it to produce a final answer. Regarding external tools, {\oursname} leverages a SQL database and a Python interpreter to process tabular data, aligning its setup with the ReAcTable framework to ensure a fair comparison. Specifically, when using Python, tables are stored as Pandas DataFrames. We do not prescribe specific tools in our framework design, making it fully flexible to use alternative tools such as R or other Python packages and data structures for table processing.

\vspace{2mm}
\noindent\textbf{Few-shot prompts for in-context learning.} For each benchmark, we construct few-shot prompts based on TQA samples in the training sets to initiate in-context learning. The few-shot prompts for each benchmark are static; that is, we use the same few-shot prompt throughout all experiments for a given benchmark and do not tune it to favor any particular instantiation of {\oursname}. Following prior LLM-based TQA frameworks \cite{zhang2024reactable,zhu2024autotqa}, we use five training samples to construct the few-shot prompts. 
For the logic agent, the prompt guides the model to perform logical reasoning and generate an instruction on how to process the table to obtain supporting evidence for the question. For the query agent, the prompt includes examples illustrating how to generate the corresponding SQL queries or Python code upon receiving an instruction.
{
For TableBench, we follow the evaluation setup in \cite{fan2024autoprep} and adopt the few-shot prompt used for the WikiTQ benchmark to evaluate the generalization ability of different TQA frameworks.
}

\vspace{2mm}
\noindent\textbf{LLMs.} We evaluate {\oursname} using eight popular open-weight LLMs. Three of them are Qwen2.5 models \cite{qwen2024qwen25technicalreport} with different sizes: 7, 14, and 72 billion parameters, referred to as Qwen2.5-7B/14B/72B, respectively. Two are Llama3.1 models \cite{dubey2024llama}: Llama3.1-8B and 70B. Additionally, we evaluate {\oursname} on a Mistral LLM \cite{jiang2023mistral}: Mistral-7B, a Gemma2 LLM: \cite{team2024gemma} Gemma2-9B, and DeepSeek-V3 \cite{liu2024deepseek}, a large open-weight LLM with 671 billion parameters. 
In addition, we also evaluate specialized coder LLMs such as Qwen-Coder-14B in our ablation studies in Section \ref{subsec:different-llm}.

\subsection{Overall Performance}\label{subsec:exp-overall}

\noindent\textbf{Evaluation results on the WikiTQ benchmark.} Table \ref{tb:overll-wiki} reports the evaluation results on the WikiTQA benchmark, from which we make the following three main observations.

First, across all LLMs, {\oursname} consistently and significantly outperforms both the CoT baseline and the state-of-the-art ReAcTable framework. For small- to medium-sized LLMs  
{\oursname} achieves a substantial improvement in test accuracy compared with ReAcTable, with a performance gap exceeding $10\%$. For large-scale LLMs, such as Llama3.1-70B, Qwen2.5-72B, and DeepSeek-V3, while ReAcTable already achieve strong performances of $65\%$, $67\%$, $70\%$, respectively, {\oursname} further enhances TQA performance, reaching a high-quality test accuracy of over $75\%$.

Second, ReAcTable sometimes causes small LLMs to underperform relative to the CoT baseline, which directly prompts the model to output the final answer without invoking external tools. For instance, when paired with Mistral‑7B and Llama3.1‑8B, ReAcTable produced incoherent responses, leading to poor TQA performance with test accuracies below $4\%$. In contrast, {\oursname} substantially improved reliability, achieving $45.9\%$ with Mistral‑7B and $64.9\%$ with Llama3.1-8B. This is probably due to the fact that ReAcTable relies on a single LLM to handle complex TQA reasoning, requiring long and intricate in‑context prompts that small models struggle to follow. By decomposing the reasoning process and simplifying the context for each agent, {\oursname} enables even small‑sized LLMs to perform effectively. We also note that newer small LLMs, such as Qwen2.5‑7B, perform much better with ReAcTable than with the CoT baseline, suggesting that recent post‑training improvements (e.g., SFT and RLHF) have strengthened their instruction‑following and in‑context learning abilities.

Third, when paired with large-scale models such as Qwen2.5-72B or Llama3.1-70B, {\oursname} surpasses the best previously reported result of $75.3\%$ \cite{zhu2024autotqa}, which was achieved by AutoTQA using GPT-4. Moreover, when equipped with DeepSeek-V3, one of the most advanced open-weight LLMs, {\oursname} sets a new state-of-the-art performance with a test accuracy of $75.8\%$. This indicates that medium- to large-sized open-weight LLMs have significant potential for handling TQA tasks with high-quality performance.

\begin{table}[!t]
\caption{Comparison of TQA performance on WikiTQ.}
\label{tb:overll-wiki}
\centering
\begin{footnotesize}
\vspace{-1mm}
\centering
\begin{tabular}{|c|cc|c|}
\hline
\textbf{Backbone LLMs} & CoT \cite{wei2022chain} & ReAcTable \cite{zhang2024reactable} & \textbf{{Ours}} \\ \hline\hline
Mistral-7B & 11.9\% & 3.6\% &  \textbf{45.9\%} \\ \hline
Gemma2-9B & 43.6\% & 46.0\% &  \textbf{64.2\%} \\ \hline
Llama3.1-8B  & 40.3\% & 2.5\% & \textbf{64.9\%} \\ 
Llama3.1-70B & 59.7\% & 65.7\% &  \textbf{75.5\%} \\ \hline
Qwen2.5-7B  & 48.7\% & 57.9\% & \textbf{68.6\%} \\ 
Qwen2.5-14B & 54.1\% & 60.7\% &  \textbf{72.1\%} \\ 
Qwen2.5-72B & 60.5\% & 67.4\% &  \textbf{75.4\%} \\ \hline
DeepSeek-V3 & 63.5\% & 70.5\% & \textbf{75.8\%} \\ \hline
\end{tabular}
\end{footnotesize}
\vspace{-1mm}
\end{table}

\vspace{2mm}
\noindent\textbf{Evaluation results on the TabFact benchmark.} Table \ref{tb:overll-tabfact} presents the evaluation results on the TabFact benchmark. Similar to the WikiTQ benchmark, the results indicate that when paired with smaller-scale LLMs such as Mistral-7B and Llama3.1-8B, the TQA performance of the ReAcTable framework falls short of practical requirements. In contrast, Orchestra demonstrates strong overall performance across all evaluated LLMs. In addition, when using the same backbone LLM, {\oursname} consistently and significantly outperforms both the CoT baseline and the ReAcTable framework. Moreover, when paired with Llama3.1-70B, Qwen2.5-72B, and DeepSeek-V3, {\oursname} achieves a high test accuracy of over $90\%$, surpassing the best prior results of $88.7\%$ achieved by AutoTQA equipped with GPT-4-turbo.

\begin{table}[!t]
\caption{Comparison of TQA performance on TabFact.}
\label{tb:overll-tabfact}
\centering
\begin{footnotesize}
\vspace{-1mm}
\begin{tabular}{|c|cc|c|}
\hline
\textbf{Backbone LLMs} & \text{CoT} \cite{wei2022chain} & ReAcTable \cite{zhang2024reactable} & \textbf{{Ours}} \\ \hline\hline
Mistral-7B & 63.2\% & 5.9\% &  \textbf{72.3\%} \\ \hline
Gemma2-9B & 70.8\% & 72.2\% &  \textbf{78.7\%} \\ \hline
Llama3.1-8B  & 69.5\% & 6.7\% &  \textbf{75.3\%} \\ 
Llama3.1-70B & 81.2\% & 82.2\% &  \textbf{90.0\%} \\ \hline
Qwen2.5-7B  & 71.1\% & 74.2\% & \textbf{82.7\%} \\
Qwen2.5-14B & 73.7\% & 79.9\% & {\textbf{88.2\%}} \\ 
Qwen2.5-72B & 82.3\% & 88.5\% & \textbf{91.1\%} \\ \hline
DeepSeek-V3 & 87.4\% & 88.9\% & \textbf{93.0\%} \\ \hline
\end{tabular}
\end{footnotesize}
\vspace{-1mm}
\end{table}

{
\vspace{2mm}
\noindent\textbf{Evaluation results on the TableBench dataset.} Table~\ref{tb:overall-tabbench} summarizes the performance of different methods on the TableBench benchmark. Compared to TabFact and WikiTQ, TableBench contains more diverse tables and more complex reasoning questions. Despite this increased difficulty, {\oursname} consistently outperforms both CoT and ReAcTable across all backbone LLMs. We observe a similar trend as in the WikiTQ and TabFact benchmarks: while ReAcTable struggles with smaller-scale models such as Mistral-7B and Llama3.1-8B, {\oursname} maintains strong performance even under these constrained settings. These results highlight the robustness and generalizability of our framework to more realistic and challenging TQA tasks, further validating its practical value for real-world data management scenarios.
}

\begin{table}[!t]
\caption{Comparison of TQA performance on TableBench.}
\label{tb:overall-tabbench}
\centering
\begin{footnotesize}
\vspace{-1mm}
\centering
\begin{tabular}{|c|cc|c|}
\hline
\textbf{Backbone LLMs} & CoT \cite{wei2022chain} & ReAcTable \cite{zhang2024reactable} & \textbf{{Ours}} \\ \hline\hline
Mistral-7B & 5.9\% & 4.0\% & \textbf{30.4\%} \\ \hline
Gemma2-9B & 34.5\% & 42.2\% &  \textbf{55.3\%} \\ \hline
Llama3.1-8B  & 33.7\% & 3.2\% & \textbf{54.6\%} \\ 
Llama3.1-70B & 55.4\% & 58.9\% & \textbf{66.1\%} \\ \hline
Qwen2.5-7B  & 31.6\% & 35.5\% & \textbf{52.6\%} \\ 
Qwen2.5-14B & 53.9\% & 55.1\% &  \textbf{59.3\%} \\ 
Qwen2.5-72B & 60.1\% & 63.0\% &  \textbf{68.1\%} \\ \hline
DeepSeek-V3 & 64.3\% & 65.3\% & \textbf{70.0\%} \\ \hline
\end{tabular}
\end{footnotesize}
\vspace{-1mm}
\end{table}

\subsection{Cost Comparison}\label{subsec:exp-cost-cmp}
We compare the computational cost of different TQA frameworks. Specifically, we set up a local API for Qwen2.5-7B-Instruct model, and measured (i) input and output token consumption, (ii) wall-clock inference time, and (iii) number of API requests per TQA question on the TableBench dataset. Specifically, we deployed the model on a machine with an NVIDIA A100-SXM4-80GB GPU, using TensorRT-LLM inference engine~\cite{tensorRT-LLM}. To ensure consistency in our measurements, we restricted deployment to a single model instance, and processed all API requests sequentially. This configuration maintained GPU memory usage at approximately 18-20 GB across all TQA methods.

The results are presented in Table \ref{tb:cmp-cost}. 
As expected, the proposed solution Orchestra incurs higher latency and token usage due to its multi-agent design and iterative reasoning. Compared to its competitors, Orchestra's additional latency is moderate, and it is compensated for by the substantial performance accuracy gains Orchestra achieves. 
Meanwhile, the end-to-end wall clock time of Orchestra remains within a practical range, i.e., tens of seconds per query, which is viable when the user interacts with the system.
Note that in applications such as financial or scientific report analysis, clinical decision support, and compliance auditing, accuracy is usually prioritized over real-time response, since users are willing to trade a few extra seconds of computation for more reliable answers. 

\begin{table}[!t]
\caption{Cost Comparison on TableBench Dataset (Per Question)}
\label{tb:cmp-cost}
\centering
\begin{footnotesize}
\vspace{-1mm}
\centering
\begin{tabular}{lrrr}
\hline

{Metrics} & CoT & ReAcTable & Ours \\ \hline
Time (sec.) & 4.6 & 19.6 & 37.3  \\
\#. API Requests & 1.0 & 10.7 & 23.7  \\ 
\#. Input Tokens & 818 & 44,243 &  179,752 \\ 
\#. Output Tokens & 292 & 1,138 & 1,641 \\\hline

\end{tabular}
\end{footnotesize}
\vspace{-1mm}
\end{table}

\subsection{Ablation Study}\label{subsec:exp-ablation}
\begin{figure}[!t]
\centering
\begin{tabular}{c}
\hspace{-2mm}\subfigcapskip=-1mm\subfigure[Evaluation results on WikiTQ]{\includegraphics[width=1.0\linewidth]{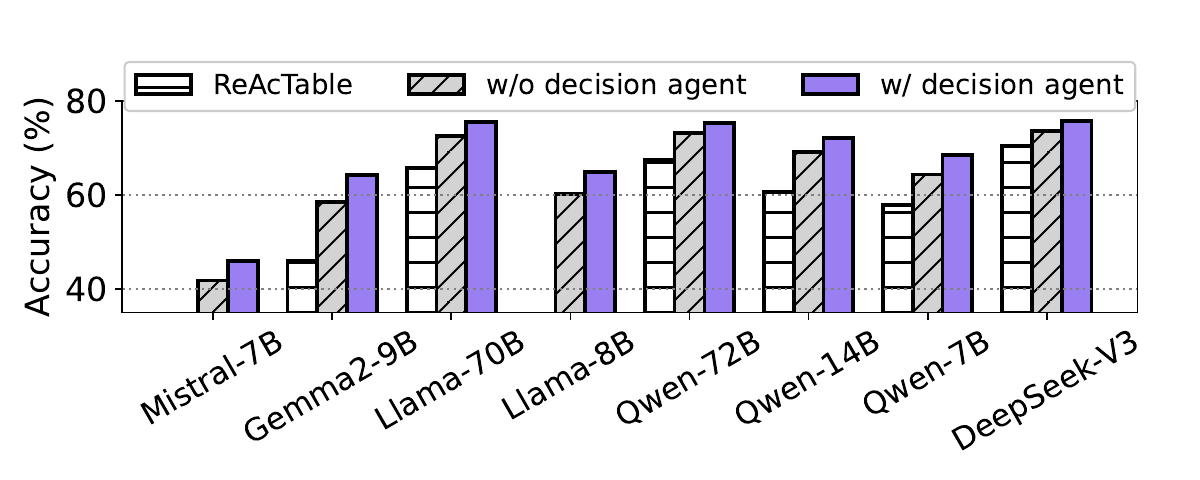}\label{subfig:ablation-wiki}} \vspace{-3mm}\\
\hspace{-2mm}\subfigcapskip=-1mm\subfigure[Evaluation results on TabFact]{\includegraphics[width=1.0\linewidth]{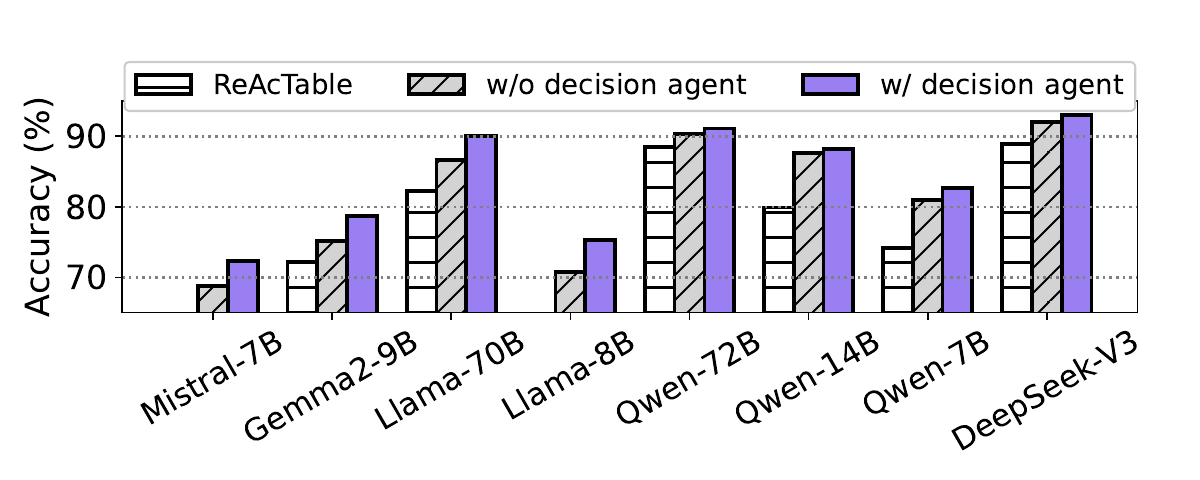}\label{subfig:ablation-tabfact}}
\end{tabular}
\vspace{-1mm}
\caption{Ablation study of reasoning context. The results of ReAcTable paired with Mistral-7B and Llama-8B are truncated by the y-axis.}
\label{fig:ablation-decision-agent}
\vspace{-1mm}
\end{figure}

We validate a key insight behind {\oursname}: refining the reasoning context can enhance TQA performance. To this end, we compare three LLM-powered solutions. The first solution, ReAcTable, serves as the baseline solution. Its reasoning context is relatively cluttered -- before making the final decision, the LLM is presented with few-shot prompts, intermediate code snippets (e.g., SQL queries and Python code) at each iteration, and all intermediate tables. The second solution is derived by removing the decision agent from {\oursname}, resulting in the two-agent system described in Section \ref{subsec:two-agent}. In this setup, the logic agent focuses solely on reasoning over intermediate tables without generating code. The performance gap between this variant and ReAcTable highlights the benefits of refining the reasoning context by eliminating code generation. The third solution is the complete {\oursname}, comprising a logic agent, a query agent, and a decision agent. Here, the decision agent produces the final answer based on a further refined reasoning context -- specifically, by filtering out the few-shot prompts. As shown in Figure \ref{fig:ablation-decision-agent}, the two-agent system outperforms ReAcTable across all models, and the complete {\oursname} achieves further gains. These results indicate that gradually refining the reasoning context leads to incremental improvements, empirically confirming the effectiveness of this approach across a wide range of open-weight LLMs. 

\begin{figure}[!t]
\centering
\begin{tabular}{cc}
\hspace{-3mm}\subfigcapskip=-0mm\subfigure[{WikiTQ}]{\includegraphics[width=0.48\linewidth]{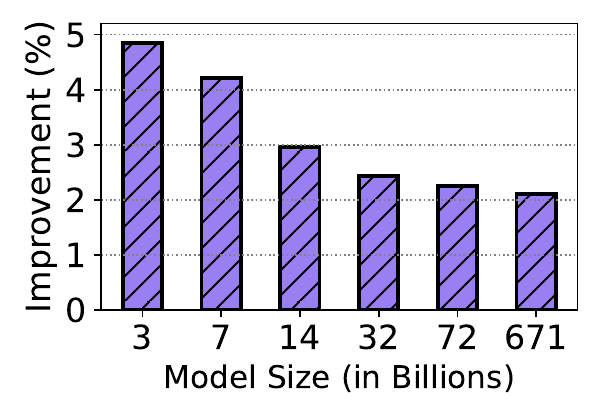}} &
\hspace{-2mm}\subfigcapskip=-0mm\subfigure[{TabFact}]{\includegraphics[width=0.48\linewidth]{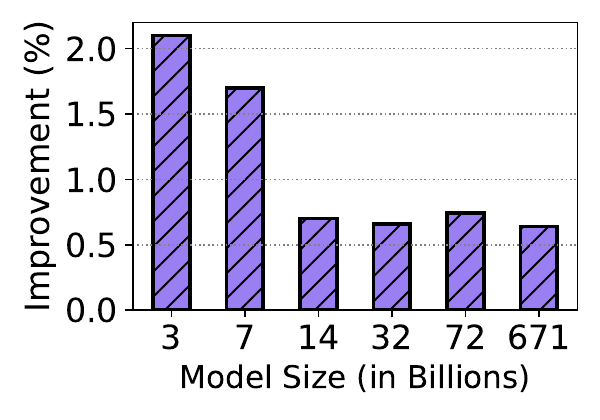}}
\end{tabular}
\vspace{-1mm}
\caption{{Effect of the decision agent on TQA accuracy.}}
\label{fig:varying-llm-size-improve}
\vspace{-1mm}
\end{figure}

{To further quantify the contribution of the decision agent, we report in Figure~\ref{fig:varying-llm-size-improve} the performance improvements achieved when it is included. We conduct this comparison using models from the Qwen2.5 family, ranging from 3B to 72B parameters, as well as the DeepSeek-V3 model with 671B parameters. 
Across all model sizes and datasets, we observe consistent accuracy gains when the decision agent is included. 
On the other hand, we also observe smaller returns of the decision agent as model size increases; this is expected, since larger models typically already exhibit stronger reasoning and longer-context capabilities, as the reviewer mentioned. Nonetheless, even for large models such as DeepSeek-V3 671B, the decision agent helps improve the accuracy of the proposed solution Orchestra.

Next, evaluate the effectiveness of the final objective calibration described in Section \ref{subsec:calibrate-obj}, which realigns the overall objective. We vary the number of Monte Carlo samples from 2 to 5. Note that using a sample size of 2 essentially bypasses the Monte Carlo method --- if a tie occurs between the two candidate answers, the first candidate is chosen by default. As shown in Figure \ref{fig:varying-mc-sample}, applying Monte Carlo calibration consistently enhances TQA performance of {\oursname} across all LLMs, empirically demonstrating the effectiveness of the Monte Carlo sampling process, and validating our insights on aligning the TQA objective in Section \ref{subsec:calibrate-obj}. 

Finally, we analyze the impact of the number of interactions between the logic and query agents. We vary the maximum number of interactions from 1 to 5, with the corresponding TQA performance reported in Figure \ref{fig:varying-iters}. When the number of interactions is minimal (e.g., 1), TQA performance drops markedly because this restricts the system to a single interaction with the table, limiting dynamic analysis and the extraction of valuable insights. As we increase the maximum number of interactions, TQA performance improves correspondingly, underscoring the importance of sufficient rounds of interaction between the logic and query agents to achieve high-quality TQA outcomes.

\begin{figure}[!ht]
\centering
\begin{tabular}{cc}
\multicolumn{2}{c}{\hspace{-4mm} \includegraphics[height=8.7mm]{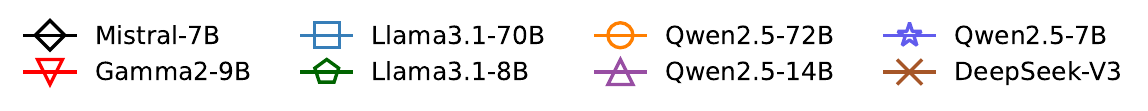}}\vspace{-4mm}\\
\hspace{-3mm}\subfigcapskip=-0mm\subfigure[WikiTQ]{\includegraphics[width=0.49\linewidth]{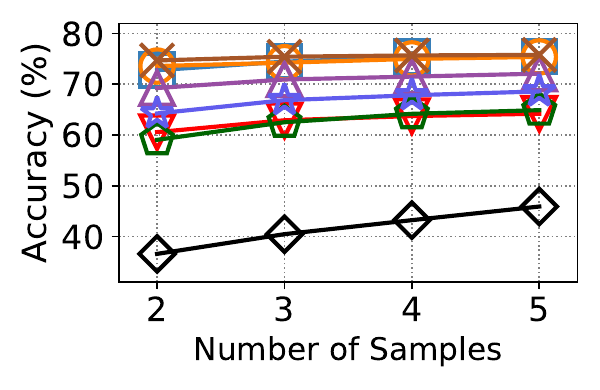}\label{subfig:vary-vote-wiki}} &
\hspace{-2mm}\subfigcapskip=-0mm\subfigure[TabFact]{\includegraphics[width=0.49\linewidth]{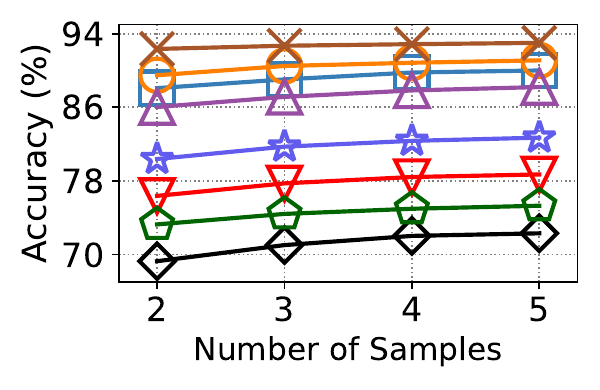}\label{subfig:vary-vote-tabfact}}
\end{tabular}
\vspace{-1mm}
\caption{Ablation study of the Monte Carlo method.}
\label{fig:varying-mc-sample}
\vspace{-1mm}
\end{figure}

\begin{figure}[!ht]
\centering
\begin{tabular}{cc}
\multicolumn{2}{c}{\hspace{-4mm} \includegraphics[height=8.7mm]{figures/consistency_legend}}\vspace{-4mm}\\
\hspace{-3mm}\subfigcapskip=-0mm\subfigure[WikiTQ]{\includegraphics[width=0.48\linewidth]{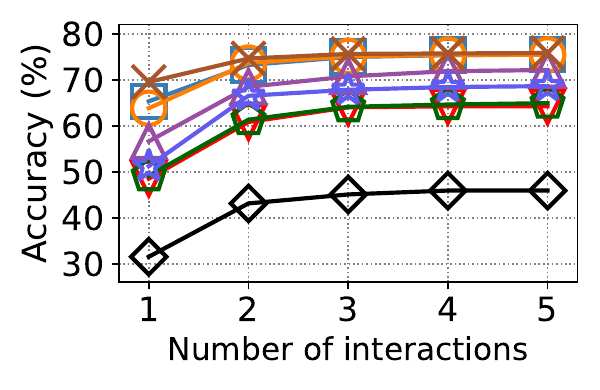}\label{subfig:vary-iters-wiki}} &
\hspace{-2mm}\subfigcapskip=-0mm\subfigure[TabFact]{\includegraphics[width=0.48\linewidth]{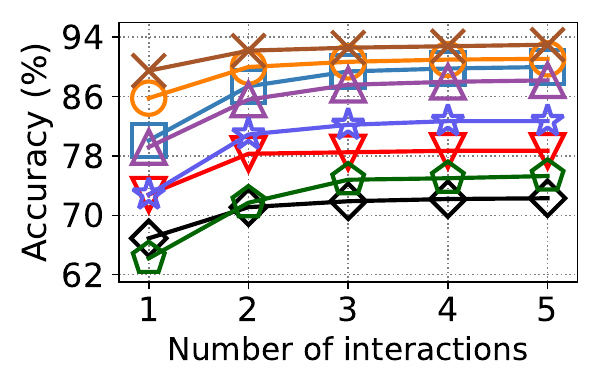}\label{subfig:vary-iters-tabfact}}
\end{tabular}
\vspace{-1mm}
\caption{Varying number of interaction rounds.}
\label{fig:varying-iters}
\vspace{-1mm}
\end{figure}
}

\subsection{Effect of Different LLMs}\label{subsec:different-llm}
\begin{figure}[!t]
\centering
\begin{tabular}{cc}
\multicolumn{2}{c}{\hspace{10mm} \includegraphics[height=12mm]{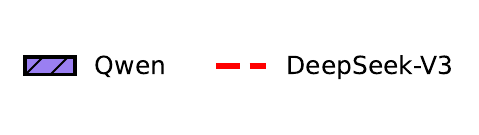}}\vspace{-8mm}\\
\hspace{-2mm}\subfigcapskip=-0mm\subfigure[WikiTQ]{\includegraphics[width=0.48\linewidth]{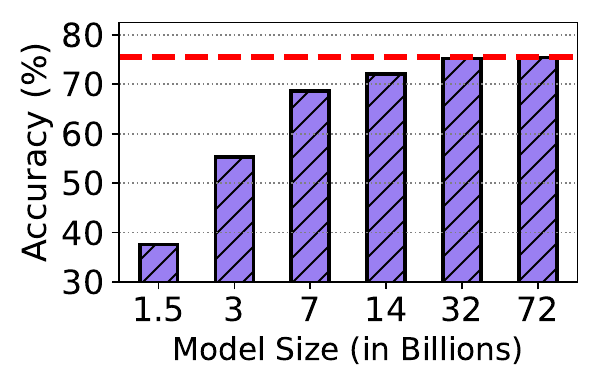}\label{subfig:vary-size-wiki}} &
\hspace{-1mm}\subfigcapskip=-0mm\subfigure[TabFact]{\includegraphics[width=0.48\linewidth]{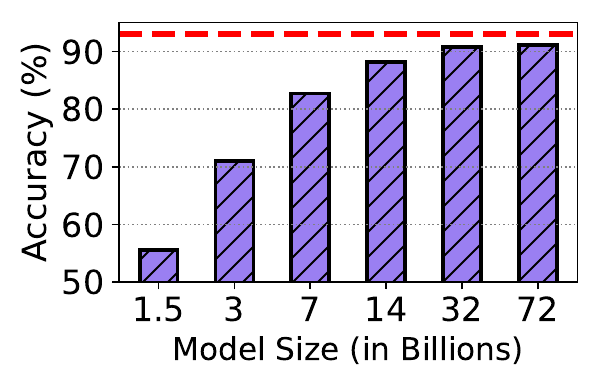}\label{subfig:vary-size-tabfact}}
\end{tabular}
\vspace{-2mm}
\caption{Effect of model size.}
\label{fig:varying-llm-size}
\vspace{-2mm}
\end{figure}

We evaluate the impact of model size by instantiating {\oursname} with various models from the Qwen2.5 series, whose sizes range from 1.5 to 72 billion parameters. As illustrated in Figure \ref{fig:varying-llm-size}, increasing the LLM size generally improves TQA performance, but with diminishing returns. For instance, expanding the model from 1.5 billion to 7 billion parameters yields significant improvements. However, further increasing the model size, such as moving from 32 billion to 72 billion parameters, or even employing models exceeding 600 billion parameters (e.g., DeepSeek-V3), result in only marginal gains. 

Moreover, we observe that when {\oursname} is paired with edge-side models such as Qwen2.5-1.5B, the {\oursname} framework fails to achieve desirable TQA performance. This is because the smaller LLMs often struggle to capture behavior patterns through in-context few-shot learning, causing the agents to deviate from their intended execution patterns. 
{
That said, we note that identifying this lower bound is itself an important insight, as it offers practical guidance: for general-purpose TQA, it is inadvisable to instantiate frameworks with existing edge-side LLMs such as Qwen2.5-1.5B. This observation also motivates future research into methods that can sustain strong TQA performance even with edge-deployable backbone LLMs. One promising direction is to construct high-quality, TQA-specific datasets and apply targeted post-training (e.g., supervised fine-tuning and RLHF) on small open-weight LLMs. This could enhance both their instruction-following ability and their robustness in generating reliable SQL/Python code for TQA tasks.
}

\balance
\section{Related Work}
\label{sec:related-work}
Large language models (LLMs) have demonstrated significant potential in data management applications \cite{fernandez2023large,li2024table,feuer2024archetype,dargahi2024dtt,luoma2025snails,omar2025dialogue,giannakouris2025lambda,yan2024gidcl,li2025llm}.
A substantial amount of recent work has explored LLM-based natural language-to-code techniques for processing table data \cite{fan2024combining,trummer2024generating,gao2024text,ren2024purple,fan2024metasql,zheng2024adapting,zhang2025clear,fan2025grounding,li2025aidsql}. These methods typically prompt an LLM to translate natural language queries into one-shot, static executable code (e.g., SQL queries or Python code) to interact with tabular data and respond to TQA queries. Although effective in some cases, previous studies \cite{chengbinding,li2024table,wangtablechain} have shown that such static approaches often struggle with TQA tasks that require multiple reasoning steps.

Another line of research has focused on enabling LLMs to process table data directly \cite{trummer2023can,ye2023large,zhang2024tablellama,li2024table,xing2024table}. Specifically, recent studies \cite{zhang2024tablellama,li2024table,xing2024table} have fine-tuned LLMs on extensive tabular datasets to improve their ability to interpret and manipulate tables. However, these solutions typically demand substantial computational resources and often lack flexibility, as they are tailored to specific TQA scenarios and struggle to generalize \cite{xing2024table}. Moreover, instead of reasoning through problems from scratch, these fine-tuned models tend to memorize patterns from the training data, which can lead to inconsistent or inaccurate responses, especially for tasks demanding precise numerical analysis or deep table understanding \cite{ji2023survey,trummer2023can,li2024table}. Consequently, fine-tuned LLMs have not yet reached state-of-the-art performance across diverse TQA tasks \cite{li2024table,xing2024table,zhang2024tablellama}.

To address these limitations, recent studies \cite{wangtablechain,zhang2024reactable} have adopted the ReAct paradigm \cite{yaoreact}, enabling LLMs to iteratively reason and interact with tables. In these frameworks, the LLM incrementally performs logical reasoning and invokes external tools to extract insights from the table, thereby improving the accuracy and reliability of the final answer. However, these methods depend on a single LLM to handle both logical reasoning and tool interaction, typically guided by complex, handcrafted few-shot prompts. As a result, these methods fail to achieve high-quality TQA performance when paired with small-sized LLMs, as illustrated in Section \ref{subsec:exp-overall}.

Recent advancements suggest that LLMs equipped with memory modules and external tools can emulate human-like interactions and decision-making processes \cite{park2023generative,wu2023autogen,shinn2024reflexion,gao2024agentscope,hong2024metagpt}.
In the context of TQA, recent studies \cite{zhu2024autotqa,fan2024autoprep} have explored multi-agent frameworks. AutoTQA \cite{zhu2024autotqa} targets scenarios involving multiple tables in specific industry settings. Specifically, they employ a planning agent that generates a static plan for interacting with various tables based on a user query, followed by sequential execution by other LLM agents. {Both the approach and the target TQA setting are fundamentally different from our {\oursname} framework, which relies on dynamic, iterative reasoning through interactions between the logic agent and the query agent to tackle TQA with single table.} This flexible, step-by-step process enables our framework to effectively handle complex TQA questions that require multi-step reasoning. %
{Further, the design of Orchestra is modular, which can be potentially extended to multi‑table scenarios by incorporating a schema‑linking/planning component, along with prompt engineering and join handling, to identify and combine relevant tables, and pass the result to the TQA workflow. This is an exciting direction for future work.} 
Another recent work, AutoPrep \cite{fan2024autoprep}, employs a multi-agent framework to address TQA in a single-table setting. However, AutoPrep focuses on designing a LLM-based data preparation process specifically tailored for TQA tasks, which is orthogonal to our goal of developing an end-to-end LLM-powered TQA framework. Investigating how to integrate AutoPrep’s data preparation techniques with our {\oursname} approach could be a promising direction for further enhancing TQA performance.

\section{Conclusion, Limitations, and Future Work}\label{sec:conclusion}
In this paper, we introduce {\oursname}, an LLM-powered multi-agent framework designed for TQA tasks. Our framework carefully orchestrates multiple LLM agents to dynamically construct reasoning chains for TQA tasks. 
By ensuring that each agent operates within a logically coherent reasoning path, our approach effectively improves the overall quality of the LLM's output. Experimental results demonstrate that {\oursname} consistently and significantly improve the performance of a wide range of LLMs across various TQA tasks.

While {\oursname} significantly improves TQA performance with open-weight LLMs, it has two limitations in specific use cases. 
First, it requires a relatively long inference time, reflecting the scaling law for LLM test-time computation \cite{kaplan2020scaling,snell2024scaling,chen2024simple}. Specifically, the framework involves multiple interactions between agents and external tools, where each reasoning step depends on previous computations. As a result, parallelization is not feasible, limiting its efficiency. This makes {\oursname} less suitable for real-time applications where low-latency responses are critical. %
Second, {\oursname} is primarily designed for handling TQA tasks based on a single table. Some applications may require answering questions based on multiple tables from different sources \cite{zhu2024autotqa}. Extending {\oursname} to support multi-table reasoning is an interesting and promising direction for future research.

\balance

\clearpage

\bibliographystyle{./bibliography/IEEEtran}
\bibliography{./bibliography/IEEEexample}

\end{document}